\documentclass[aps,prl,superscriptaddress,twocolumn]{revtex4-1}
\pdfoutput=1

\usepackage{graphicx}

\newcommand{\fig}[1]{Fig.~\ref{#1}}

\newcommand{\bigref}[1]{Ref.~\onlinecite{#1}}

\begin{document}

\title{Antisite traps and metastable defects in Cu(In,Ga)Se$\rm_2$ thin-film solar cells \\studied by screened-exchange hybrid density functional theory}

\author{Johan Pohl}
\affiliation{Institut f{\"u}r Materialwissenschaft, Technische Universit{\"a}t Darmstadt, Petersenstr. 32, D-64287 Darmstadt, Germany}
\author{Thomas Unold}
\affiliation{Helmholtz-Zentrum Berlin, Hahn-Meitner-Platz 1, D-14109 Berlin, Germany}
\author{Karsten Albe}
\affiliation{Institut f{\"u}r Materialwissenschaft, Technische Universit{\"a}t Darmstadt, Petersenstr. 32, D-64287 Darmstadt, Germany}

\date{\today}

\begin{abstract}
Electronic structure calculations within screened-exchange hybrid density functional theory show that $\rm Cu_{In,Ga}^0$ antisites in both CuInSe$\rm_2$ and CuGaSe$\rm _2$ are localized hole traps, which can be attributed to the experimentally observed N2 level. In contrast, $\rm Ga_{Cu}$ antisites and their defect complexes with copper vacancies exhibit an electron trap level, which can limit the open-circuit voltage and efficiency in Ga-rich Cu(In,Ga)Se$_2$ alloys. Low-temperature photoluminescence measurements in CuGaSe$_2$ thin-film solar cells show a free-to-bound transition at an energy of 1.48 eV,   in very good agreement with the calculated transition energy for the $\rm Ga_{Cu}$ antisite. Since the intrinsic DX center $\rm In_{DX}$ does not exhibit a pinning level within the band gap of CuInSe$\rm _2$, metastable DX behaviour can only be expected for $\rm Ga_{Cu}$ antisites.
\end{abstract}

\maketitle

 Thin-film solar cells with aborbers based on Cu(In,Ga)Se$_2$ alloys currenly achieve record efficiencies of 20.3\% and represent one of the most promising technology for large-scale industrial production \cite{JHL+11}. Efficiencies up to 14.2\% can also be obained with the ternary boundary phase CuInSe$\rm_2$, whereas the efficiencies of pure CuGaSe$\rm _2$ cells are still limited to below 10 \% \cite{CKC+07}. Intrinsic point defects acting as recombination centers are likely to limit the open-circuit voltage and therefore the solar cell efficiency of these absorbers. The point defect physics of these chalcopyrites has been extensively studied by experimental methods, such as electrical and optical spectroscopy methods, and theoretical approaches, mostly calculations based on density functional theory \cite{ZWZ97,ZWZ+98,WZZ98,ZPL+04,PZL+05,LZ05,LZ06,LZ08,PA10,PKA11,OGS+11}. A hole trap level in the range between 0.15-0.35~eV, often named N2, has been observed using admittance and deep-level transient spectroscopy (DLTS) and Hall measurements in CuInSe$\rm _2$ \cite{WHM+96,HJR+00,MPR07}, CuGaSe$\rm _2$ \cite{JRN+00,SR08,MPR07,HJR+00,KIZ+11} and Cu(In,Ga)Se$\rm _2$ \cite{HIS98,SHB+98,HCS04,WHM+96,HJR+00}). A measured activation energy between 0.05-0.20~eV, has been attributed to an interface defect in Cu(In,Ga)Se$\rm_2$ and was named N1 \cite{HIS98}, although this denomination is ambiguously used and it is unclear wether the N1 response is associated to a defect at all \cite{ECN+11}. Indeed, various metastable effects have been oberserved in CIGSe devices such as persistent photoconductivity \cite{MEP+02}, the increase of the open-circuit voltage upon white-light soaking \cite{RR87}, an increase of the space-charge upon illumination \cite{IS96} or reverse-biasing \cite{HRS+99} accompanied with a decrease of the fill factor \cite{IBS+03} as well as capacitance relaxation on long time scales after light-soaking \cite{ESM+98}.

Based on electronic structure calculations within local density functional theory, two intrinsic point defects and their complexes with copper vacancies have been proposed to exhibit metastable properties in CuInSe$_2$ and CuGaSe$_2$: the intrinsic indium and gallium DX centers $\rm(In,Ga)_{DX}$\cite{LZ08}\footnote{Intrinsic DX centers in CIGS can be regarded as $\rm (In,Ga)_{Cu}$ antisites displaced to a three-fold coordinated off-lattice site, which may then act as recombination centers because they exhibit defect levels within the gap \cite{LZ08}.} and the selenium vacancy $\rm{V_{Se}}$ \cite{LZ05} or selenium vacancy--copper vacancy complex $\rm{V_{Se}-V_{Cu}}$ \cite{LZ06}. These theoretical results have been invoked\cite{CIZ+08,UI09,IUE09,SIP+10} to explain the experimentally observed light and voltage-bias induced metastabilities. However, metastable point defects are not the only possible explanation. Copper migration in the space charge region\cite{BEG+97,GKC+00}, deep acceptor levels in the CdS buffer layer\cite{EGS+98}, an electron-injection dependent barrier at the molybdenum back-contact of the device \cite{ ECN+11}, or the presence of a p+-layer in conjunction with a shallow donor level at the buffer absorber interface\cite{NBH+98} have all been put forward as possible explanations. Therefore, metastabilities in Cu(In,Ga)Se$_2$ based devices and their possible relation to the N2 and N1 levels remain puzzling and it is not clear whether a single explanation is sufficient to explain all of the observed phenomena \cite{ECN+11,Siebentritt11}. Despite extensive efforts, no concise picture of the point defect physics that matches with all of the experimental findings has yet emerged.

In this letter, we show that $\rm Cu_{In,Ga}^0$ antisites are localized hole traps in both CuInSe$\rm_2$ and CuGaSe$\rm _2$, which can be attributed to the experimentally observed N2 level, while $\rm Ga_{Cu}$ antisites and their defect complexes with copper vacancies exhibit an electron trap level, which can limit the open-circuit voltage and efficiency in Ga-rich Cu(In,Ga)Se$_2$ alloys. Our results are based on electronic structure calculations within screened-exchange hybrid density functional theory and are put in context to low-temperature photoluminescence measurements of CuGaSe$_2$ thin-film solar cells.

\begin{figure}
\centering
\includegraphics[width=60mm]{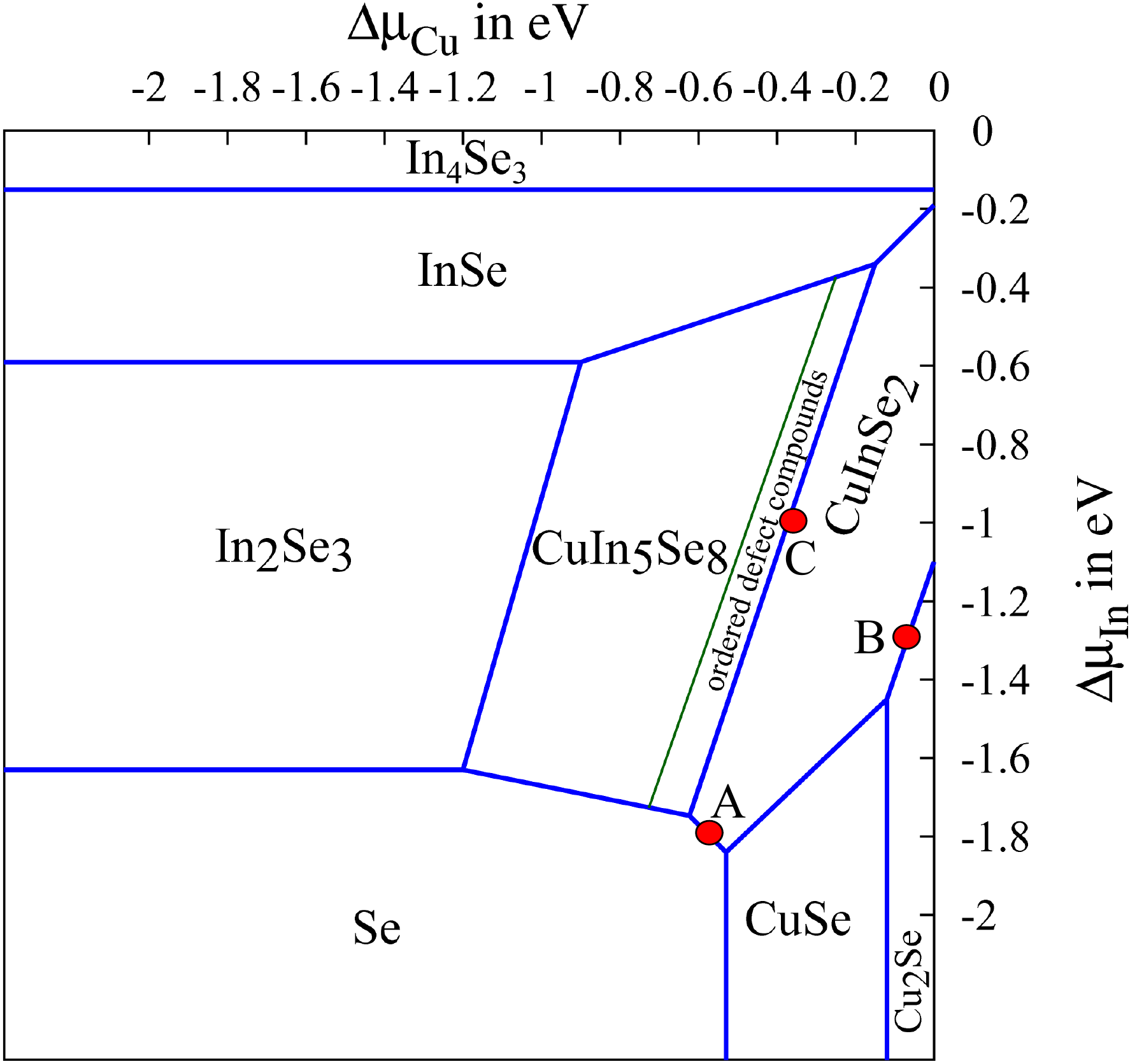}
\caption{Stability diagram for CuInSe$\rm_2$. The defect formation enthalpies in Fig. \ref{fig_formation} are discussed in terms of the chemical potential at points A,B and C. The stability range for CuGaSe$\rm_2$ (not shown) has the same shape, but a somewhat larger extent due to its higher formation enthalpy (CuInSe$\rm _2$: -2.37 eV, CuGaSe$\rm_2$: -2.67 eV). }
\label{fig_stability}
\centering
\end{figure}

We have carried out hybrid density functional calculations using the HSE06 functional \cite{HSE03,HSE06} as implemented in VASP\cite{VASP} with an adapted exchange-screening parameter of 0.13 $\rm \AA^{-1}$\footnote{The standard value for the fraction of exact exchange of 0.25 and a plane-wave energy cutoff of 350 eV was used.}, which allows to closely match the experimentally observed band gaps for both CuInSe$\rm{_2}$, CuGaSe$\rm{_2}$ and other chalcopyrite phases\cite{PA10}. The approach allows to overcome the band-gap problem, to directly analyze defect levels within the gap and  improves the description of localized and correlated copper $d$ electrons \cite{LAD+11, ZYS+11}. Thus, using the HSE06 functional will lead to more accurate formation enthalpies of point defects than LDA calculations. A 2x2x2 $\rm \Gamma$-centered k-point grid has been used for supercells with 64 atoms as well as with 216 atoms. Ion positions were relaxed until forces were converged to below 0.05 and 0.1 eV/\AA, for supercells of 64 and 216 atoms, respectively. The calculation of supercells of 216 atoms with a 2x2x2 k-point grid are computationally extremely costly, but necessary in order to observe unambiguously localized defect levels of $\rm Cu_{In,Ga}^0$ and $\rm Ga_{Cu}^0$. All reference phases presented in the stability diagram (\fig{fig_stability}) were calculated using the same functional. The point defect formation enthalpies were calculated as function of the chemical potentials of the constituents $\Delta \mu_{i}$ referenced to the elemental phases and the Fermi energy $\epsilon_F$ according to the common formula as e.g. in Ref. \cite{PZL+05}.  The potential alignment and image charge corrections have been carefully carried out as described in \bigref{LZ08a}.

\begin{figure}
\centering
\includegraphics[width=\columnwidth]{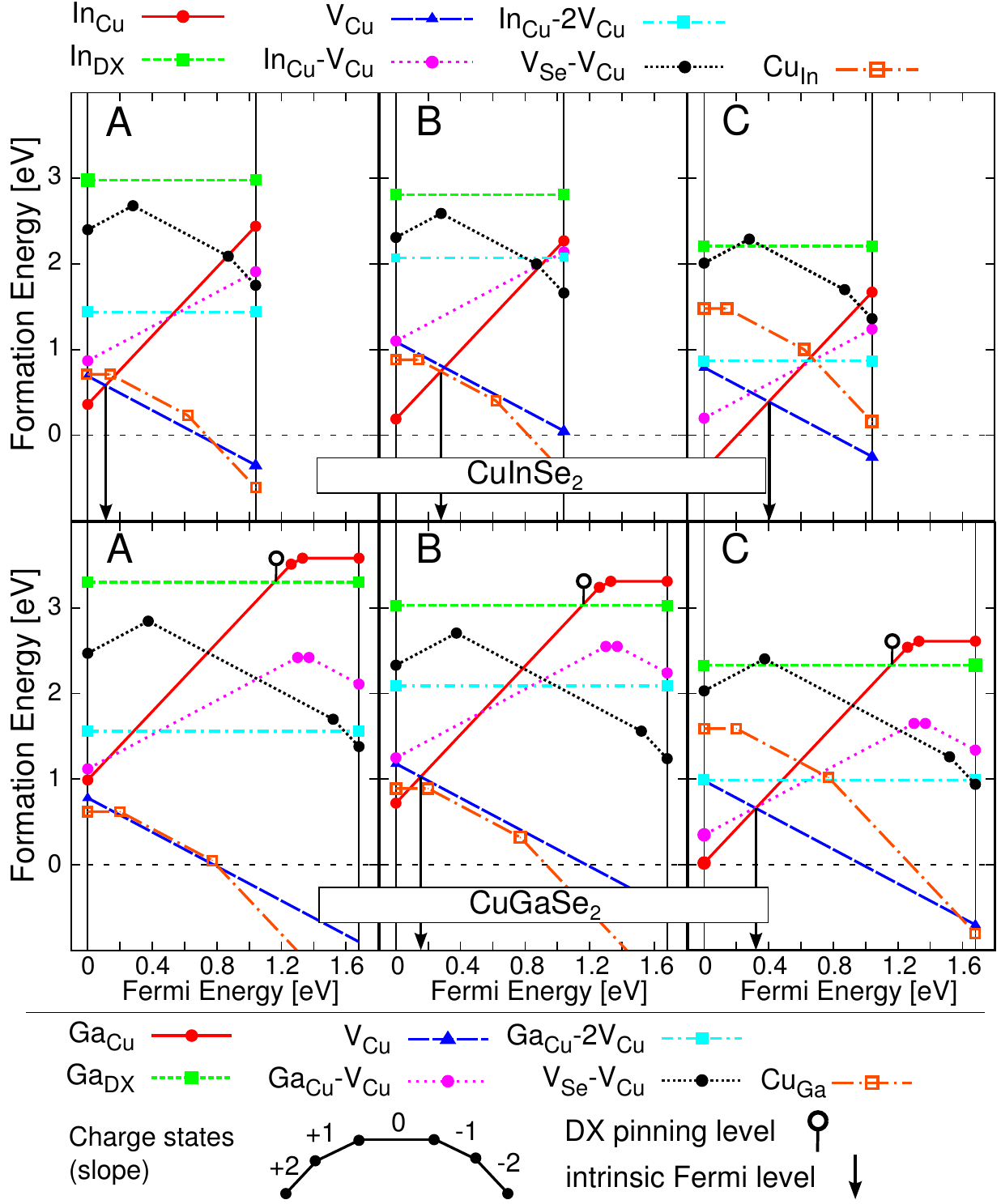}
\caption{Defect formation enthalpies, charge transition levels and the determined intrinsic Fermi levels in CuInSe$\rm_2$ and CuGaSe$\rm _2$ for chemical potentials corresponding to points A,B and C in Fig. \ref{fig_stability}.}
\label{fig_formation}
\centering
\end{figure}

 Since high-efficiency Cu(In,Ga)Se$\rm _2$ absorber material is prepared under a highly selenium-rich atmosphere, e.g. with a selenium to metal flux ratio of 5 \cite{HML+03}, it is instructive to interpret the defect physics for this material under selenium-rich conditions on the Se-Cu(In,Ga)Se$\rm _2$ phase boundary in the calculated stability diagram (point A in \fig{fig_stability}).  In contrast to a previously calculated stability diagram \cite{PZL+05}, our calculations show a phase boundary between Cu$\rm_2$Se and CuInSe$\rm_2$, which is in line with the experimental phase diagram and the observation of Cu$\rm_2$Se precipitates under certain processing conditions\cite{HGE+98} (point B). For Cu-poor compositions, which yield the highest conversion efficiency, Cu(In,Ga)Se$\rm_2$ is a highly compensated semiconductor. In this case, the charge neutrality condition and thus the Fermi-energy is essentially determined by the concentration of donors and acceptors with the lowest formation energies (see Fig. \ref{fig_formation}). For the chemical potentials at points A, B and C the material turns out to be p-type, while it becomes n-type for maximal Cu- and In-rich conditions (not shown). Fig. \ref{fig_formation} shows the calculated defect formation enthalpies for the various chemical potentials.

One of the most intriguing result is that $\rm Cu_{In,Ga}$ antisites in both materials, CuInSe$\rm_2$ and CuGaSe$\rm _2$, can have equally low formation enthalpies as copper vacancies and thus also act as compensating defects. This finding is consistent with large concentrations of $\rm Cu_{In}$ in $\rm CuInSe_2$ recently reported using wavelength dispersive x-ray diffraction even for copper-poor material \cite{SST+11}. However, when the chemical potentials are shifted towards metal-rich conditions (e.g. point C), it is seen that the formation enthalpy of  $\rm Cu_{In,Ga}$ at the intrinsic Fermi level (vertical arrow) increases, while it does not change much for $\rm {(In,Ga)}_{Cu}$ and $\rm V_{Cu}$. 

\begin{figure}
\centering
\includegraphics[width=\columnwidth]{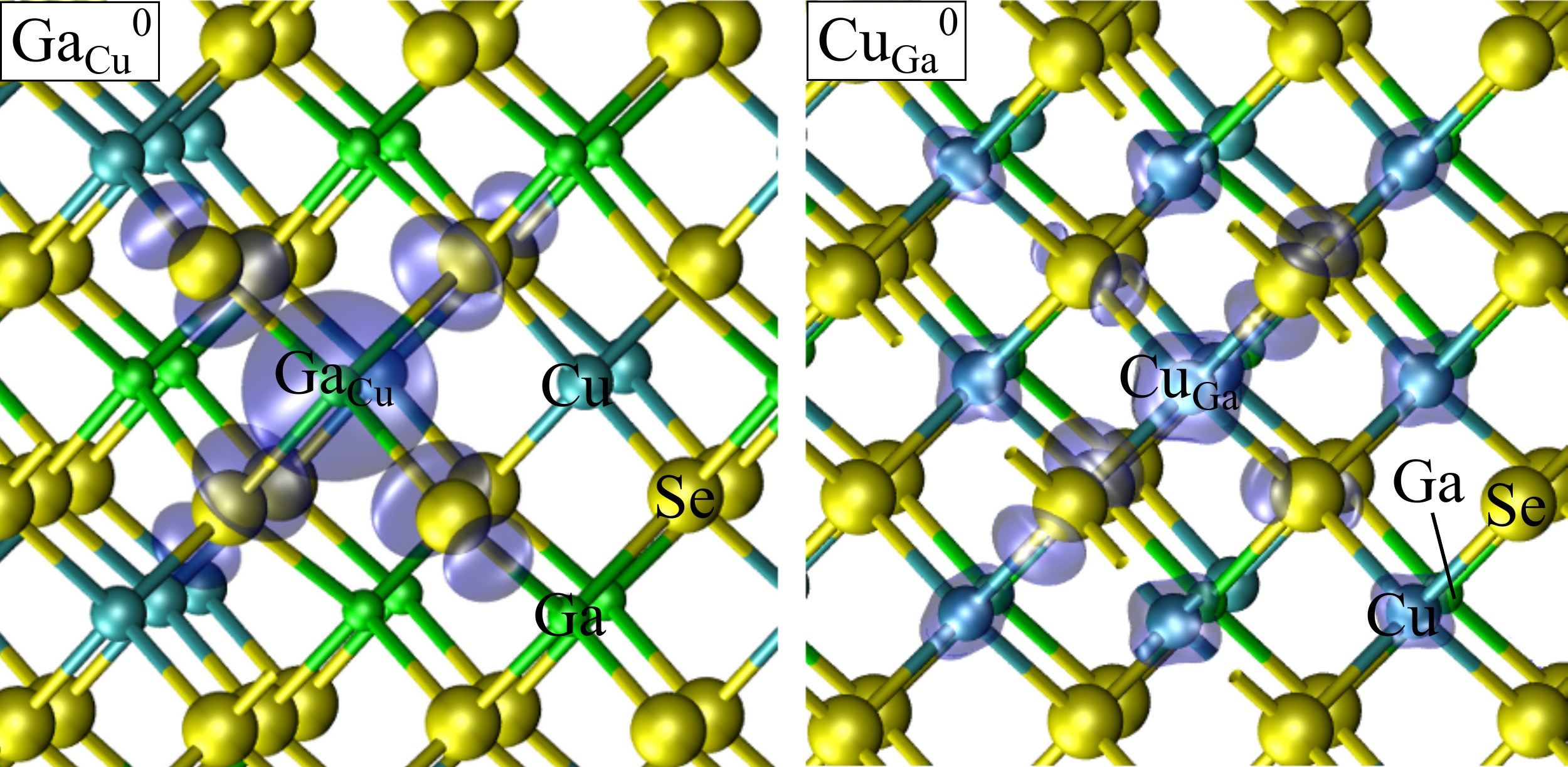}
\caption{Defect level electron charge density of $\rm Ga_{Cu}^0$ (left: isosurface 0.03 $e\rm \AA^{-3})$ and hole density of $\rm Cu_{Ga}^0$ (right: isosurface 0.02 $e\rm\AA^{-3}$) as obtained in supercells with 216 atoms.}
\label{fig_charge}
\centering
\end{figure}

The analysis of the density of states of $\rm Cu_{In,Ga}^0$ and $\rm In_{Cu,Ga}^0$ reveals an empty narrow defect band above the VBM for $\rm Cu_{In}$ in CuInSe$\rm_2$ (at 0.27 eV) and $\rm Cu_{Ga}$ in CuGaSe$\rm_2$ (at 0.32 eV), which can trap two holes (the hole density of the empty single-particle defect state of $\rm Cu_{Ga}$ as obtained from the calculation is displayed in \fig{fig_charge}). Furthermore, a localized electron trap level emerges for $\rm Ga_{Cu}^0$ at a single-particle energy of 1.17 eV above the VBM in CuGaSe$\rm _2$ (see also Fig. \ref{fig_charge}). Forming defect complexes such as  $\rm Ga_{Cu}-V_{Cu}^-$ and $\rm (Ga_{Cu}-2V_{Cu})^{-2}$, does not affect the position of the defect level. The defect level of $\rm In_{Cu}^0$ is found to be resonant within the CB for CuInSe$\rm _2$ at 1.48 eV and for CuGaSe$\rm _2$ at 1.46 eV approximately independent of gallium content. All individual single-particle defect levels are approximately constant on an absolute energy scale, when comparing their position in CuInSe$\rm_2$ and CuGaSe$\rm_2$ relative to the respective VBM. However, the defect levels of $\rm In_{Cu}^0$ and $\rm Ga_{Cu}^0$ as individual defects do not align on an absolute scale, the $\rm In_{Cu}^0$ defect level being 0.29-0.41 eV higher than the one of $\rm Ga_{Cu}^0$ in both CuInSe$\rm_2$ and CuGaSe$\rm_2$ as host material. 

\begin{figure}
\centering
\includegraphics[width=\columnwidth]{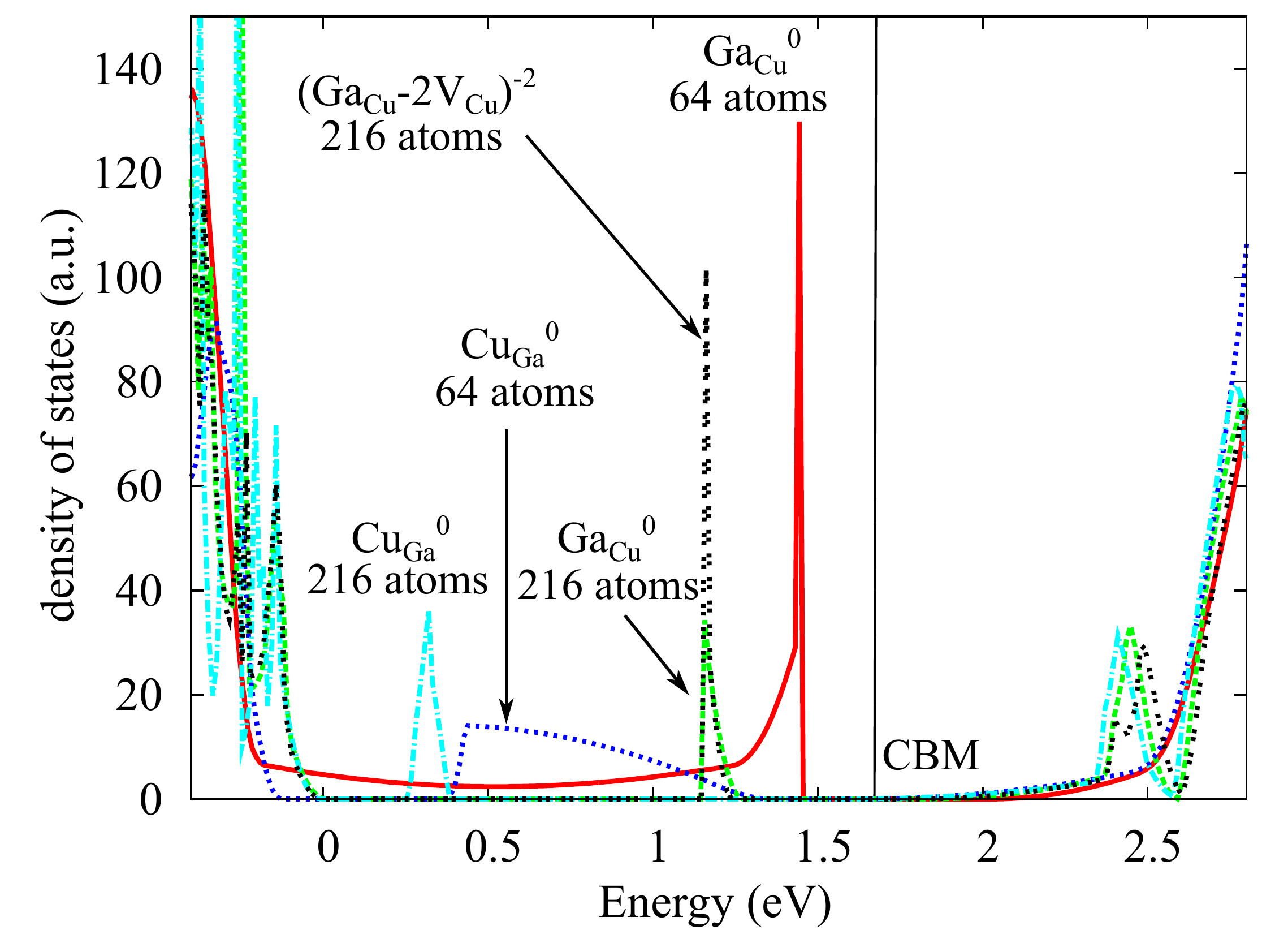}
\caption{Aligned density of states of $\rm Cu{Ga}^0$, $\rm Ga_{Cu}^0$ and $\rm (Ga_{Cu}^0-2V_{Cu})^{-2}$ in CuGaSe$\rm _2$ in supercells of 64 and 216 atoms as obtained with the adapted HSE06 functional.}
\label{fig_DOS}
\centering
\end{figure}

Here it should be noted that defect calculations on Cu(In,Ga)Se$_2$ have so far been performed with supercells smaller or equal to 64 atoms. However, very disperse defect bands appear within the gap for the cells with 64 atoms (see Fig. \ref{fig_DOS}), which indicates significant self-overlap of the defect wavefunctions. If local or semilocal density functionals are used, delocalization also occurs in large supercells up to 216 atom, which is why localized deep hole traps have not been identified in the past. This proves that an accurate nonlocal treatment of exchange and correlation and large supercells are crucial for obtaining the correct localization behaviour of the $\rm Cu_{In,Ga}^0$ defect.

The $\rm Cu_{In,Ga}$ antisites localize holes, are abundant under typical preparation conditions (up to $\rm 10^{20} cm^{-3}$ at 850 K deposition temperature and the calculated thermal transition energy for the process $\rm Cu_{Ga}^-+h^+_{VB} \rightarrow Cu_{Ga}^{0}$ of 0.20 eV  agrees with experimental measurements (e.g. 0.1-0.3 eV in Ref. \cite{HCS04} and 0.1-0.2 eV in Ref. \cite{SHB+98}). 
 Given these evidences it is safe to conclude that the N2 hole trap level is due to the $\rm Cu_{In,Ga}$ antisite. The fact that this level does not occur in all samples can be explained with differing formation enthalpies relevant for different preparation conditions (compare points A,B,C in \fig{fig_formation}). 
 
The $\rm Ga_{Cu}^0$ antisite, in contrast, shows a clearly localized electron trap level at a single-particle energy of 1.17 eV above the VB in CuGaSe$\rm_2$ and at 1.07 eV in CuInSe$\rm_2$ very close to the CBM. Therefore this antisite defect becomes increasingly deep when Ga is alloyed into CuInSe$_2$, due to the rising CB. Since $\rm Ga_{Cu}^0$ is expected to occur in large quantities in $\rm Cu(In,Ga)Se_2$ due to its low formation enthalpy, it may limit solar cell efficiency when Ga alloying is used to increase the band-gap: When the CBM is raised above the position of the $\rm Ga_{Cu}$ defect level, recombination through this defect may limit the open-circuit voltage. Since the associates of  $\rm Ga_{Cu}$ with copper vacancies display the same single-particle defect level as the non-complexed antisite, the complexes may cause the same limitations. With regard to the defect complexes $\rm (In,Ga)_{Cu}-2V_{Cu}^0$ we find that their formation enthalpies are not particularly low (Fig. \ref{fig_formation}). In fact, the total binding energy of the complexes with respect to the isolated charged species is only about 0.3 eV.

\begin{figure}
\centering
\includegraphics[width=\columnwidth]{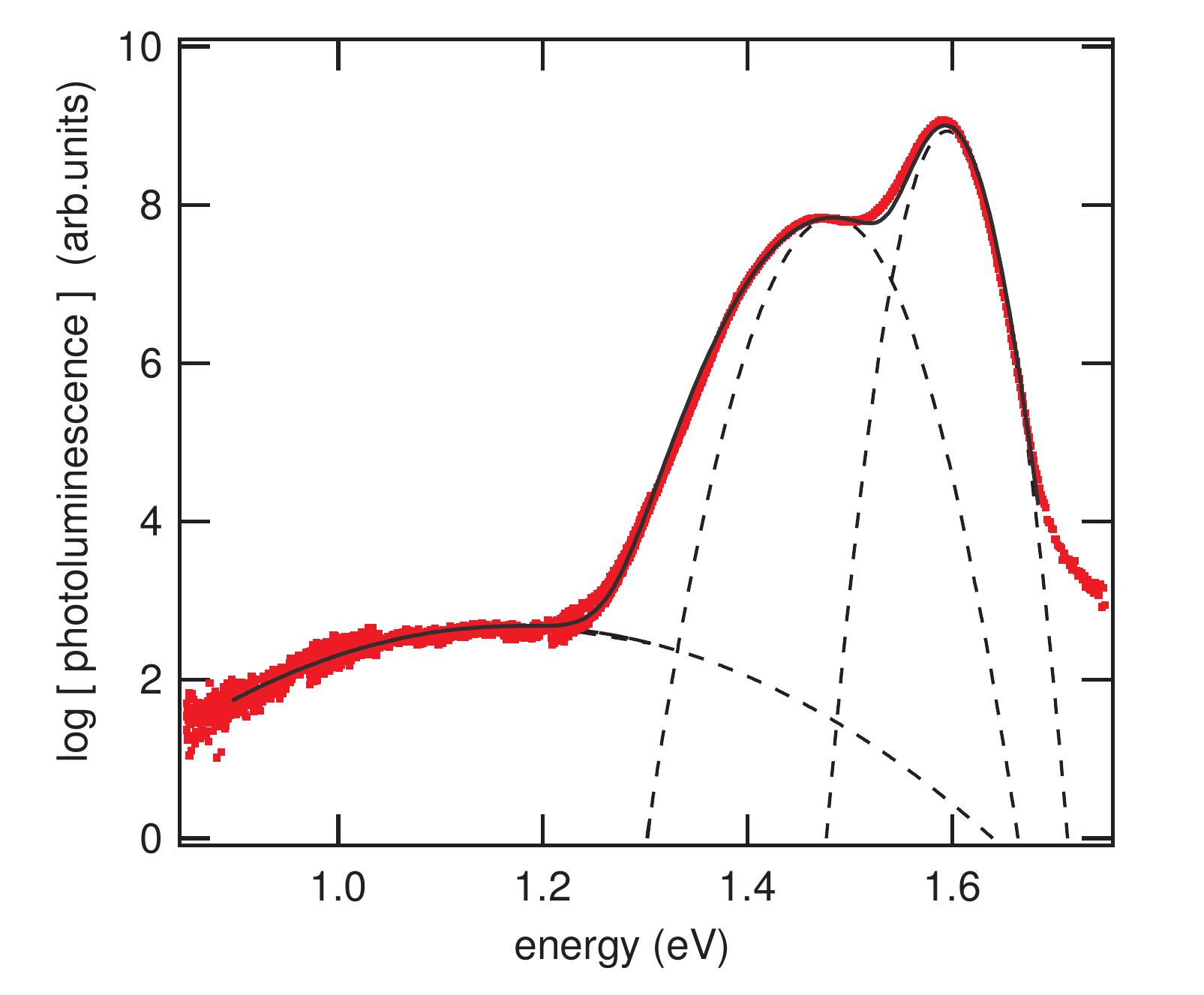}
\caption{Photoluminescence spectrum at T=15 K measured for CuGaSe$_2$ thin films. The black solid line indicates a fit consisting of three gaussian-shaped transitions at E1 = 1.6 eV, E2 = 1.48 eV and E3 = 1.17 eV (dashed lines).}
\label{fig_PL}
\centering
\end{figure}

It might be tempting to relate the experimentally observed N1 signature to the normal configuration of the $\rm Ga_{Cu}$ defect in $\rm Cu(In,Ga)Se_2$. However, we note that such an assignment is contradicted by the observation of N1 in pure $\rm CuInSe_2$ and the fact that its activation energy does not change with increasing Ga-content \cite{EUC+09,TKR01}.

To confirm our theoretical results for the $\rm Ga_{Cu}$ defect, we conducted temperature-dependent photoluminescence measurements on a CuGaSe$\rm_2$ thin-film solar cells prepared by a three-stage coevaporation process, as used for high-efficiency chalcopyrite solar cell devices\cite{CKC+07}. For the thin film absorber investigated, a ratio [Cu]/[Ga] = 0.87 was measured by x-ray fluorescence analysis and the accompanying solar cell showed a device efficiency of 7\%. Photoluminescence (PL) was measured using a 670 nm diode-laser as excitation source and a thermoelectrically-cooled InGaAs array coupled to a 0.5 m spectrograph for luminescence detection, with the sample placed in a closed-cycle helium cryostat. A photoluminescence spectrum obtained at 15 K is shown in Fig. \ref{fig_PL}. Three recombination peaks located at 1.6 eV, 1.48 eV and 1.17 eV can be clearly distinguished. The temperature and excitation intensity dependence of these peaks is consistent with an assignment of the peak at 1.6 eV to a tail-to-band transition, commonly observed in Cu-poor chalcopyrites \cite{KCY+99}, and the assignment of the peak at 1.48 eV to a free-to-bound transition. The latter transition energy is in excellent agreement with the calculated optical emission energy for the recombination process $\rm Ga_{Cu}^++h^+_{VB} \rightarrow Ga_{Cu}^{++} $ in $\rm CuGaSe_2$ (1.44 eV) \footnote{Since Koopman's theorem does not apply in density functional theory, the single-particle energies may differ from absorption and photoluminescence energies.}. The temperature and excitation dependence of the third broad peak observed at about 1.17 eV is consistent with a donor-acceptor transition. A possible candidate for this transition is the process $\rm Ga_{Cu}^++Cu_{Ga}^0 \rightarrow Ga_{Cu}^{++} + Cu_{Ga}^{-1}$, i.e. the recombination of a single electron localized on a $\rm Ga_{Cu}$ antisite with a neighbouring $\rm Cu_{Ga}$ hole trap by radiative tunneling, which has a calculated transition energy of 1.02 eV.

For metastabilities originating from intrinsic DX centers to occur, it is necessary that a DX pinning level exists within the gap\cite{LZ08}. From our results (Fig. \ref{fig_formation}), we conclude that such a pinning level only occurs for $\rm Ga_{Cu}$ antisites in CuGaSe$_2$ ($E_{\rm DX,pin}^{\rm CuGaSe_2} = \rm 1.16\;eV$), but not for $\rm In_{Cu}$ antisites in CuInSe$_2$ ($E_{\rm DX,pin}^{\rm\;CuInSe_2} = \rm 1.31\;eV$, well above the CB). Therefore, Fermi-level pinning and metastable effects due to intrinsic DX centers may occur in larger band gap Cu(In,Ga)Se$_2$ materials, but not for the ternary CuInSe$_2$. In order to assess the energy differences responsible for the different DX pinning levels as compared to Ref. \cite{LZ08}, it is instructive to compare the \it uncorrected \rm formation enthalpies of $\rm In_{Cu}^{2+}$ and $\rm In_{DX}$ in CuInSe$_2$ within our approach to the ones obtained from LDA applying only the static +U valence band correction of Ref. \cite{LZ08}, which we were able to reproduce. The uncorrected formation enthalpy of $\rm In_{Cu}^{2+}$ within our approach is 0.4 eV lower, while the one of $\rm In_{DX}$ is 0.32 eV higher. These energy differences, which result from the different treatments of exchange and correlation within HSE06 as compared to the LDA, which result mostly from the improved description of the Cu $d$ electrons, directly cause the change in the DX pinning levels.

We have also investigated $\rm V_{Se}-V_{Cu}$ vacancy pairs, which have been held responsible for a variety of metastability phenomena in Cu(In,Ga)Se$_2$. With respect to this defect pair our hybrid functional calculations yield comparable charge transition levels (+/-) and metastable relaxation behavior as previously found in LDA-based calculations \cite{LZ06}. However, the formation enthalpies of this defect complex are higher than 2 eV at the relevant Fermi levels at points A,B and C in Fig. \ref{fig_formation}). Thus, under thermal equilibrium conditions, $\rm V_{Se}-V_{Cu}$ associates should occur only in minor quantities (below $\rm 10^{12} cm^{-3}$ at 850 K deposition temperature as estimated from the formation enthalpies). Thus, these defects can only cause metastable phenomena if the material is prepared under far-from-equilibrium conditions.

In conclusion, we have shown that $\rm Cu_{In,Ga}$ defects create a localized hole trap level in both CuInSe$_2$ and CuGaSe$_2$. Given the large amount of experimental evidence for a hole trap level in the range 0.15-0.35 eV, we conclude that $\rm Cu_{In,Ga}$ defects should be assigned to the N2 level. $\rm Ga_{Cu}^0$ and its complexes display a localized defect level within the gap in CuGaSe$\rm _2$ and CuInSe$\rm _2$. These defects are thus a likely cause for the limited efficiency of CuGaSe$_2$ based wide band  devices. Since $\rm In^0_{DX}$ does not exhibit a Fermi pinning level in CuInSe$\rm _2$, metastable DX behaviour can only be expected for $\rm Ga_{Cu}$ antisites.

We acknowledge useful discussions with Dr. P\'eter \'Agoston and Prof. Andreas Klein (TUD), Dr. R. Caballero for CuGaSe$\rm_2$ sample growth, S. Kretschmar for the photoluminescence measurements, grants of computer time on JUROPA at the J\"ulich Supercomputing Center and funding within the GRACIS project (BMBF).

\bibliographystyle{apsrev}

\end{document}